# Promoted Electronic Coupling of Acoustic Phonon Modes in Doped Semimetallic MoTe$_2$

Xiangyue Cui[†], Hejin Yan[†], Xuefei Yan[§¶], Kun Zhou[‡], Yongqing Cai[†]*

[†]*Joint Key Laboratory of the Ministry of Education, Institute of Applied Physics and Materials Engineering, University of Macau, Taipa, Macau, China;*

[‡]*School of Mechanical and Aerospace Engineering, Nanyang Technological University, 50 Nanyang Avenue, Singapore 639798, Singapore; Environmental Process Modelling Centre, Nanyang Environment and Water Research Institute, Nanyang Technological University, 1 Cleantech Loop, Singapore 637141, Singapore;*

[§]*School of Microelectronics Science and Technology, Sun Yat-Sen University, Zhuhai 519082, China;*

[¶]*Guangdong Provincial Key Laboratory of Optoelectronic Information Processing Chips and Systems, Sun Yat-Sen University, Zhuhai 519082, China*

*Corresponding author. Email: yongqingcai@um.edu.mo

**ABSTRACT**

As a prototype of the Weyl superconductor, layered molybdenum telluride (MoTe$_2$) encompasses two semimetallic phases (1T' and T$_d$) which differentiate from each other via a slight tilting of the out-of-plane lattice. Both phases are subjected to serious phase mixing which complicates the analysis of its origin of superconductivity. Herein, we explore the electron-phonon coupling (EPC) of the monolayer semimetallic MoTe$_2$, without phase ambiguity under this thickness limit. Apart from the hardening or softening of phonon modes, the strength of the EPC can be strongly modulated by doping. Specifically, longitudinal and out-of-plane acoustic modes are significantly activated for electron doped MoTe$_2$. This is ascribed to the presence of rich valley states and equispaced nesting bands which are dynamically populated under charge doping. Through comparing the monolayer and bilayer MoTe$_2$, the strength of EPC is found to be less likely to depend on thickness for neutral samples but clearly promoted for thinner samples with electron doping, while for hole doping, the strength alters more significantly with the thickness than doping. Our work explains the puzzling issue of the doping sensitivity of the superconductivity in semimetallic MoTe$_2$ and establishes the critical role of activating acoustic phonons in such low-dimensional materials.

## INTRODUCTION

The emergence of two-dimensional (2D) transition metal dichalcogenides (TMDs) has aroused extensive interest in recent years. They exhibit versatile properties including semiconducting, semimetallic, and topological-insulating behaviors, making it an enticing platform for multiple applications[1-3] in electronics,[4-6] optoelectronic,[7-9] spintronics[10] and valletronics.[11] Molybdenum telluride (MoTe$_2$) as one of emergent TMDs shows additional features with respect to its rich polymorphism, including exotic p-type semiconduction,[12] quantum spin Hall states (QSHSs)[13] and topological superconductivity.[3,14] Thus far, three crystalline polymorphic phases of MoTe$_2$ - 1H (hexagonal), 1T' (monoclinic) and T$_d$ (orthorhombic) phases have been found. Controllable growth of massive metallic phases was demonstrated for monolayer[15] and few-layer MoTe$_2$ via oxygen deficient MoO$_x$ ($x < 3$) precursors compared with the 1H phase from a MoO$_3$ precursor.[12] The energy difference between the hexagonal 1H- and distorted octahedral 1T'-MoTe$_2$ is tiny and only ~35 meV per formula unit,[16,17] which would be further modulated via strain and doping,[18] enabling the construction of a coplanar MoTe$_2$ heterostructure. The 1T'/1H coplanar-MoTe$_2$ quantum dot generates the Kubo gap in the seeding quantum dot of the 1T' phase, which is appealing for nanophotonics and reconfigurable devices.[18] Interestingly, for bulk MoTe$_2$, a structural phase transition from the centrosymmetric 1T' phase (the room temperature phase) to the non-centrosymmetric T$_d$ phase occurs with reducing temperature to ~250 K, and the latter phase shows a topological superconductivity[3] and is regarded as a prototype Weyl superconductor.[19] The critical temperature ($T_c$) of the superconductivity of semimetallic T$_d$-MoTe$_2$ increases dramatically with decreasing its thickness.[20]

Recent experiment showed that with increasing pressure, the temperature of the phase transition from the 1T' phase to the T$_d$ phase reduces.[3] A dome-shape of $T_c$ was observed together with the shrinkage of the van der Waals (vdW) gap[3] and modulated density of states at the Fermi level under pressure.[2] While bulk 1T' and T$_d$ phases show strikingly different topological and superconducting behaviors which can be modulated by a mechanical approach,[21] both phases share the same in-plane atomic structure in the monolayer limit which shows a novel antilocalization effect[15] and promoted $T_c$.[20] Unfortunately, mechanism of the correlation of the superconductivity with the intralayer electron-phonon coupling (EPC) in monolayer semimetallic MoTe$_2$ is less understood. In particular, without the interlayer shift and rigid layer vibrations, a good knowledge of the phonon mode-decomposed EPC in monolayer 1T' (or T$_d$) MoTe$_2$ is highly desired, which can be set as the foundation for clarifying the differences in multilayer polymorphs.

Owing to the confined and limited electronic states of 2D materials, doping via chemical adsorbates or physical gating has been proven to be an important approach for adjusting the Fermi level and (de)populate otherwise empty (occupying) levels.[22] Thus, doping can alter the population of the electronic bands which renormalizes the EPC and the superconducting states. Search for superconductors in doped 2D materials such as graphene and TMDs has attracted great and ever-increasing interest.[23-28] For MoTe$_2$, chemical doping was intentionally introduced via methods such as S-,[29] Se-,[30,31] and Re-substitutions,[32] which strongly modulate the superconductivity.[33] There is a promoted superconductivity observed through isoelectronic substitution of S/Se with Te in MoTe$_2$.[29-31] Here the enhancement would be related to the charge transfer from the dopant to the host lattice due to a different electronegativity. The lattice distortion associated with the difference in the atomic radius may also play some role in activating the EPC through exerting the strain. The native Te-vacancy in semimetallic MoTe$_2$ is responsible for a temperature-manipulated transport in a manner equivalent to electrical gating.[34] Through precisely generating Te-vacancies with inducing electron-doping, the T$_d$-MoTe$_{2-x}$ samples show a promoted superconducting transition at 2.1 K compared with stoichiometric samples (no superconducting state down to 10 mK).[35] Also, the charge doping can modulate structural stability and induce phase transitions for layered MoTe$_2$.[36] Besides that, there was a recent breakthrough achieved for synthesizing centimeter-sized high-quality bilayer and trilayer 1T'-MoTe$_2$ films successfully.[37] So far, it is still unclear how electronic doping alters the atomic vibrations and the momentum resolved electron-phonon coupling for the phonons.

In this work, via first-principles calculations, we mainly focus on the nature of the EPC in monolayer semimetallic 1T'-MoTe$_2$ and its evolution with doping, and also explore the effect of dimensionality on the activation of the EPC of acoustic modes from bilayer to bulk 1T' phase. Examining the monolayer systems avoids the complexity associated with the interlayer sliding in multilayers T$_d$ and 1T' phases. We reveal that the acoustic phonons are prone to be responsible for the coupling with the electrons. The longitudinal (LA) and out-of-plane (ZA) acoustic phonons become strongly activated under electron doping while the transversal acoustic (TA) modes barely keep silent. We attribute this sensitive change to the synergy of rich distribution of valleys together with equispaced conduction and valence bands around the Fermi level.

## METHODS

Our calculations including the structural optimizations and the interatomic force constant (IFC) for the phononic properties were performed on the basis of density functional theory (DFT)[38] as implemented in the Quantum ESPRESSO code.[39] We selected the norm-conserving pseudopotentials from the Hartwigesen-Goedecker-Hutter PP table with the local density approximation (LDA) as the exchange-correlation functional, and set the plane-wave energy cutoff at 110 Ry, along with a dense $k$-mesh of 10 × 18 × 1 Monkhorst-Pack scheme. The detailed convergence tests are given in Figure S1 and S2. The thickness of vacuum region is about 15 Å to prevent interaction between interlayers. The relaxed lattice constant under doping together with the effect of spin-orbit coupling (SOC) has been examined. We found there exists a continual and slight expansion of lattice constant (Figure S3) due to the quantum Hook's law.[40]

For the calculations of EPC parameter and $T_c$, the density functional perturbation theory (DFPT)[41] was adopted on an uniform unshifted denser $k$-mesh of 10 × 18 ×1 and interpolated to a finer $k$-mesh of 20 × 36 ×1. We chose a $q$ mesh of 5 × 9 × 1, Gaussian smearing method with 0.01 Ry broadening and a typical Coulomb pseudopotential of $\mu^* = 0.1$ to calculate the EPC. Likewise, the calculation of the bilayer 1T' phase adopted the same parameters as the monolayer case, while for bulk 1T' phase, a denser $k$-mesh of 8 × 16 × 4 was used and further interpolated to a finer 16 × 32 × 8 $k$-mesh, alongside with the $q$ mesh of 4 × 8 × 2 being applied. The effect of SOC on the EPC of monolayer 1T' structure was not considered due to computational limitations, but we examined its effect on the frequencies of optical phonon branches at Γ point (Figure S4), and found the effect is negligible, consistent in previous studies.[21,42] With regard to electronic properties, we calculated both with and without SOC along high-symmetry path, especially for band structures (Figure S5) and Fermi surfaces (Figure S6). Some electronic analysis were assisted by calculations using the Vienna *ab initio* simulation package (VASP),[43] where the generalized gradient approximation (GGA) of the Perdew-Burke-Ernzerh (PBE) exchange-correlation function[38] was used. The valence electron configurations for the Mo and Te element are $4d^55s^1$ and $5s^25p^4$, respectively. The values of the energy cutoff chosen as 600 eV and the $k$-mesh as 8 × 16 × 1 were set.

# RESULTS AND DISSCUSSION

We constructed the semimetallic monolayer MoTe$_2$ from bulk 1T'-MoTe$_2$, the pristine monolayer is characterized by a monoclinic unit cell with *P*2$_1$/*m* space group, and it adopts a centrosymmetric structure same as the parent bulk phase. The optimized lattice parameter for 2D MoTe$_2$ are *a* = 6.29 Å, *b* = 3.39 Å, and *β* = 93.92° (without SOC), which is consistent with previously reported studies,[18,33,44] where Mo atoms are located at the center of a distorted octahedron of Te atoms with each of Mo atom is surrounded by two unequivalent Te atoms.

To illustrate the semimetallicity of MoTe$_2$, we conducted the analysis of the electronic properties through projecting the atomic orbitals of Mo and Te. The projected band structure is shown in Figure 1a and the energy landscape for electronic bands crossing the Fermi level within the whole Brillouin zone is shown in Figure 1b. The metallicity of the 1T'-MoTe$_2$ is evident by the crossing Fermi level through the weak dispersive bands around the Γ point, which mainly consist of Mo-4*d* and Te-5*p* orbitals. The electron localization function (ELF) as shown in Figure 1c indicates that the degree of spatial localization is strongly around the Te atom. The weak ELF between Mo and Te atoms suggests weak covalent bonds, in line with the previous report.[14] Through Bader charge analysis, we found that each Mo atom loses 0.48 *e* and each Te atom obtains 0.24 *e*.

Two half-filled bands cross over the Fermi level with hole pockets at Γ point, labeled as "H1" and "H2" in Figure 1a and b, are responsible for the intrinsic superconductivity. Band projected charge density (Figure 1e) shows that the band with "H1" involving hybridized states of Mo-Te and Mo-Mo, reflecting a bonding-like nature. In contrast, the states in band containing "H2" are mainly concentrated around Mo atoms (Figure 1f), showing a more-or-less antibonding Mo-4*d* nature. Real-space analysis of the distribution of the two bands reflects the dominant components from the Mo orbitals, accounting for the strong coupling with those high-frequency Mo-involved modes around 196 and 251 cm$^{-1}$ as shown below. With regard to the effect of SOC, the hole pockets at Γ point are nearly spin degenerate, and the electron pocket at X point exhibits a tiny spin splitting for the neutral 1T' phase while the splitting becomes more prominent under doping (Figure S5).

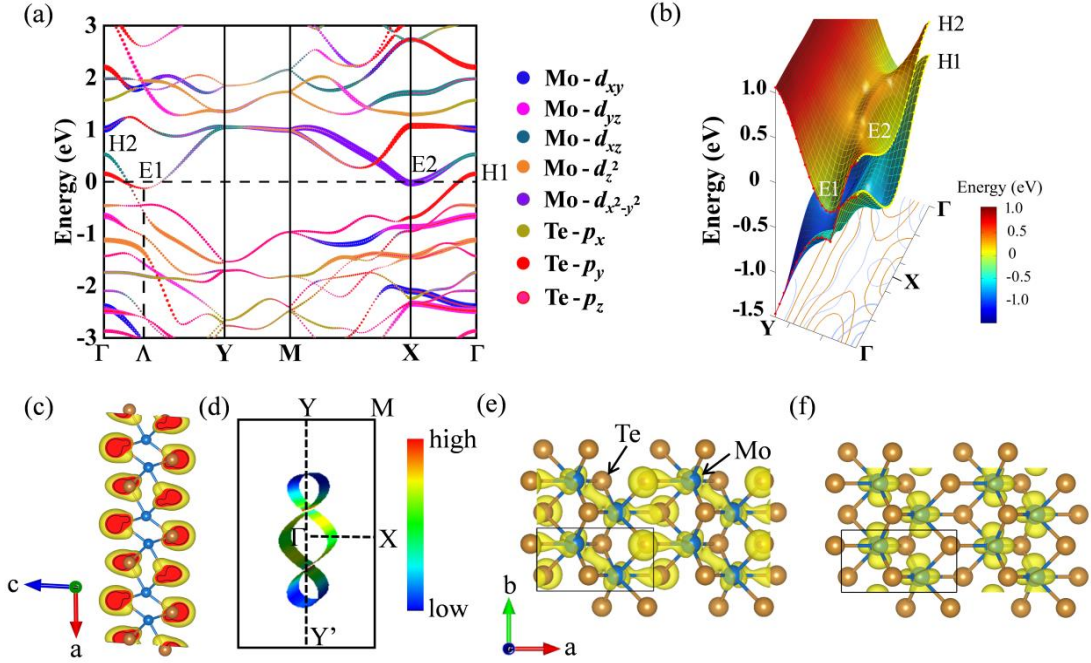

**Figure 1.** The electronic properties of pristine monolayer 1T'-MoTe$_2$. (a) Orbital-resolved electronic band structure with circles whose sizes are proportional to the weight of each orbital, where bands labeled "H1", "H2", "E1" and "E2" correspond to hole and electron pockets. (b) Energy landscape of electronic bands around Fermi level within the 2D Briullouin zone. (c) Electron localization functions (isosurface level = 0.75) and (d) corresponding Fermi surface without doping, where the colorbar on the right corresponds to Fermi velocity. The charge density distribution of two bands cross-over Fermi level containing "H1" and "H2" pockets in (e) and (f), respectively.

To estimate the potential superconductivity for pristine 1T'-MoTe$_2$, we calculated the phonon dispersion relations with the wave vector $q$ along the high-symmetry path and the corresponding electron-phonon interactions using the McMillan-Allen-Dynes equation[45] based on the Bardeen-Cooper-Schrieffer (BCS) theory,[46] which are sketched in Figure 2a. Our calculations suggest that monolayer 1T'-MoTe$_2$ is an intrinsic superconductor with a $T_c$ of 0.67 K, the EPC constant $\lambda$ = 0.40, logarithmic average of the phonon frequencies $\omega_{\log}$ = 164.52 K. The above results are derived

from[45]

$$T_c = \frac{\omega_{\log}}{1.2} \exp\left[\frac{-1.04(1+\lambda)}{\lambda - \mu^*(1+0.62\lambda)}\right] \quad (1)$$

where $\omega_{\log}$ is the logarithmic average of phonon frequency, $\lambda$ is a dimensionless physical quantity (defined in eq. 4 below) used to measure the strength of electron-phonon interaction, and $\mu^*$ represents Coulomb pseudopotential.

The Eliashberg spectral function with frequency $\omega$ defined as[47]

$$\alpha^2 F(\omega) = \frac{1}{N_0 \pi} \sum_{\mathbf{q}\nu} \frac{\gamma_{\mathbf{q}\nu}}{\omega_{\mathbf{q}\nu}} \delta(\omega - \omega_{\mathbf{q}\nu}) \quad (2)$$

where $N_0$ refers to electronic density of states at the Fermi surface, $\gamma_{\mathbf{q}\nu}$ and $\omega_{\mathbf{q}\nu}$ represent the linewidth and the frequency with phonon mode $\nu$ and momentum $\mathbf{q}$, respectively. The expression for $\gamma_{\mathbf{q}\nu}$[48] can be estimated as

$$\gamma_{\mathbf{q}\nu} = 2\pi \omega_{\mathbf{q}\nu} \sum_{mn} \int \frac{dk}{\Omega} |g_{mn}(k,q)|^2 \delta(\varepsilon_{nk} - \varepsilon_F) \delta(\varepsilon_{mk+q} - \varepsilon_F) \quad (3)$$

where $\Omega$ is the volume of the Brillouin zone, the term $g_{mn}(\vec{k},\vec{q})$ represents the EPC matrix element, which describes the probability of electron scattering by phonons with momentum $\vec{q}$, the $\varepsilon_{n\vec{k}}$ and $\varepsilon_{m\vec{k}+\vec{q}}$ are the Kohn-Sham energy, and $\varepsilon_F$ is the Fermi level. The EPC constant $\lambda$ is calculated by[45]

$$\lambda = 2 \int_0^\infty \frac{\alpha^2 F(\omega)}{\omega} d\omega \approx \frac{1}{N_q} \sum_{\mathbf{q}\nu} \lambda_{\mathbf{q}\nu} \quad (4)$$

where $N_q$ is the number of q points and $\lambda_{\mathbf{q}\nu}$ is the mode resolved electron phonon coupling.

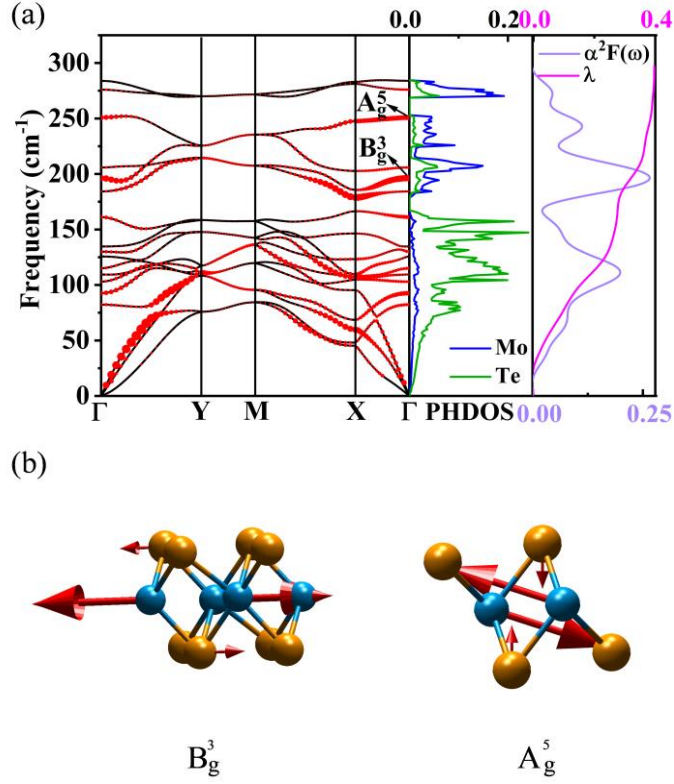

**Figure 2.** (a) Phonon band structure, projected phonon density of states (PHDOS), Eliashberg spectral function $\alpha^2F(\omega)$, and accumulative electron-phonon coupling $\lambda(\omega)$ for monolayer 1T'-MoTe$_2$. Radius of the red circles scale with the magnitude of wave vector and mode resolved $\lambda_{\mathbf{q}\nu}$. (b) The vibration patterns for two typical Raman-active optical modes - $B_g^3$ (196 cm$^{-1}$) and $A_g^5$ (251 cm$^{-1}$) at Γ point, and the lengths of arrows reflect the amplitudes of these vibrational modes, where blue and yellow spheres represent Mo and Te atoms, respectively.

Figure 2a shows the absence of imaginary frequencies in phonon dispersion, ensuring the stability of the freestanding monolayer semimetallic MoTe$_2$. The variation of the projected phonon density of states (PHDOS), cumulative EPC constant $\lambda(\omega)$ and electron-phonon spectral function $\alpha^2F(\omega)$ are also presented. The phonon spectrum can be largely grouped into three different parts: low frequency part with $\omega < 166.5$ cm$^{-1}$, intermediate part from 178.9 to 252.0 cm$^{-1}$, and high frequency part with $\omega > 269.5$ cm$^{-1}$, which are respectively dominated by vibrating the heavier Te atoms, coupled vibrations of Mo and Te atoms, and the lighter Mo atoms. According to the phonon dispersion with weighted component of $\lambda_{\mathbf{q}\nu}$ as shown in Figure 2a, the LA modes along the Γ-Y direction between 0 and 100 cm$^{-1}$ and the optical modes at Γ-point with frequencies of $B_g^3$ (196 cm$^{-1}$) and $A_g^5$ (251 cm$^{-1}$) show a strong EPC. This indicates that these phonons tend to suffer more scattering from electrons than the others.

The most predominant feature from the $\lambda_{\mathbf{q}\nu}$ projected dispersion of Figure 2a is the presence of strongly scattered flat phonon bands largely residing along Γ-X path. These less dispersive phonon branches at 251, 196 and 100 cm$^{-1}$ are related to the polar Mo-Te vibrating modes, and account for the peaks in the $\alpha^2F(\omega)$ spectrum function while their overall contributions to PHDOS are less evident. The broad width and a steady EPC for all the modes with different momenta along the X direction suggest the presence of continually scattered electronic states that have a strong coupling via these distorted polar modes.

Figure 2b shows the atomic vibrational displacements of the two optical modes with a strong $\lambda_{\mathbf{q}\nu}$ at the Γ: $B_g^3$ mode (196 cm$^{-1}$) with the collinear in-plane atomic vibrations and the $A_g^5$ mode (251 cm$^{-1}$) in which the Te atoms vibrate out of plane while the Mo atoms move largely in-plane with a slight out-of-plane tilt. For both modes, the patterns involve out-of-phase vibrations of the neighboring Mo atoms and adopt a gerade symmetry under inversion operation. We can also see that the vibrating amplitudes of Mo atoms are significantly higher than those of Te atoms. Frozen phonon method was adopted to calculate the evolution of the total energy of displaced structure along the eigenvectors of these two modes. Both modes induce apparent anharmonic behaviors (Figure S7). Interestingly, deformation potential calculation shows that the in-plane $B_g^3$ mode disturbs the electronic bands associated with the Mo-4$d$ at the Y-M path while the out-of-plane tilted $A_g^5$ mode causes significant shifts of nearly all the bands of MoTe$_2$ (Figure S8).

To identify the electronic mechanisms and contribution to EPC of phonons, we next explore the evolution of the EPC of 1T'-MoTe$_2$ with charge doping. The charge doping was simulated by varying the number of (added/depleted) valance electrons in the cell within a compensating jellium background, which allows populating/emptying the states around the Fermi level. The doping concentration $n$ is estimated by $n = x/S$ (in unit of e/cm$^2$), where $x$ is the number of doping charges and $S$ is the area of unit cell. The *n* we considered are + 4.64 × $10^{13}$ e/cm$^2$, + 9.19 × $10^{13}$, - 4.73 × $10^{13}$ and - 9.55 × $10^{13}$ e/cm$^2$, respectively. A positive value of *n* corresponds to electron doping, while negative ones represent hole doping. As shown in Figure 3a-d, except for the $n_h$ = -4.73 × $10^{13}$ e/cm$^2$ with a negligible negative frequency of -0.91 cm$^{-1}$ around the Γ as similar to borophene,[49] all the doped structures are all dynamically stable. The hole doping slightly stiffens the phonons, while electron doped 1T'-MoTe$_2$ shows a remarkable phonon softening of modes at the zone-edge X point and LA modes. We have also considered the correction of vacuum treatment associated with the 2D system as implemented in Quantum Espresso, and the updated EPC under electron doping is shown in Figure S9. Both calculations with and without this correction are largely consistent and both indicate that the acoustic modes make a dominant contribution.

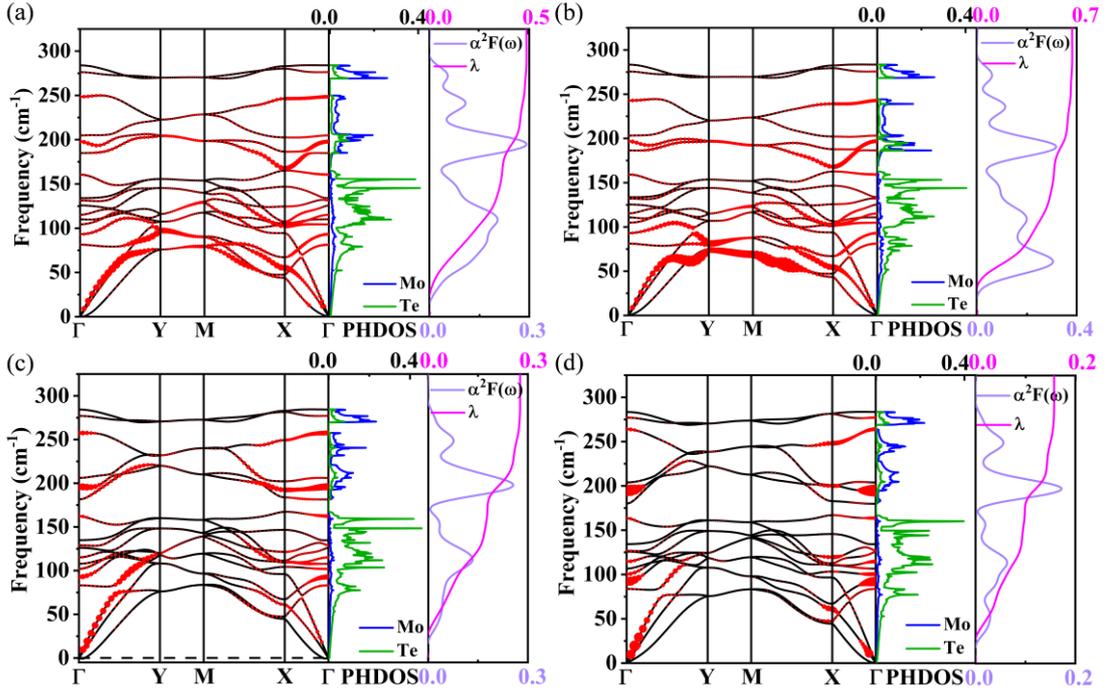

**Figure 3.** Phonon dispersions with wave vector and mode resolved $\lambda_{qv}$, projected phonon density of states, Eliashberg spectral function $\alpha^2F(\omega)$ and the electron-phonon integral, $\lambda(\omega)$, under different doping concentrations. Electron doping with (a) $n_e$ = 4.64 × $10^{13}$ e/cm$^2$ and (b) $n_e$ = 9.19 × $10^{13}$ e/cm$^2$, hole doping with (c) $n_h$ = -4.73 × $10^{13}$ e/cm$^2$ and (d) $n_h$ = -9.55 × $10^{13}$ e/cm$^2$.

The frequencies of the nine Raman-active modes at zone center under doping are listed in Table S1 together with a comparison with the experiments. Our results are consistent with experimental Raman spectrum showing five prominent peaks with $A_g^1$, $B_g^2$, $A_g^2$, $A_g^3$ and $A_g^5$ at 80, 163, 114, 128 and 252 cm$^{-1}$, respectively.[2] Our work shows the frequencies are not sensitive to charge doping, except for the $A_g^4$ and the $A_g^5$ modes softening for electron doping and the $A_g^3$ mode with showing an abnormal hardening behavior under electron doping (Figure S10).

Notably, a strong variation of the EPC of acoustic phonons with doping is found. With the increasing doping of electrons (Figure 3a and b), the EPC of LA mode along the Γ-Y and Y-M high symmetry lines, ZA along the X-M direction and TA mode around Y point are enhanced significantly. This is quite different from the neutral (Figure 2) and moderate hole doping cases as shown in Figure S11a, where only the LA along the Γ-Y is highly coupled. The broadening of these acoustic phonons under electron doping arises from the softening modes that become more strongly coupled with electronic states and thus a promoted EPC.

The effect of doping on the EPC strength of acoustic phonons is also vividly visualized in the 2D mapping of $\lambda_{\mathbf{q}v}$ with the Brillouin zone, as shown in Figure 4. Without doping (Figure 4a), only the LA branch appears an appreciable EPC with a "dome" shape aside G-Y/2 close to the long-wave length limit, while the ZA and TA branches are largely silent and have a weak EPC. Under electron doping ($n_e = 9.19 \times 10^{13}$ e/cm$^2$) as shown in Figure 4b, the region of LA with relatively appreciable EPC (i.e. $\lambda > 0.2$) significantly spreads out to cover most of the whole zone. Meanwhile the ZA mode aside the M-X within the up left corner becomes activated. Eigenvector analysis reveals the ZA mode adopts a rocking vibration of Mo (2) - Te atoms circling around the stationary Mo (1) within the z-x plane (see inset of Figure 4b). In contrast, the TA mode shows a weak enhancement at corners around X and Y points. Under hole doping with $n_h = -9.55 \times 10^{13}$ e/cm$^2$ (Figure S11a), the LA branch still dominates the EPC albeit TA and ZA shows an enhanced contribution.

As the thickness increases, the overall $\lambda$ increases slightly from 0.40 (monolayer) to 0.46 (bilayer) for 1T' MoTe$_2$. However, the distribution and mapping of the wave vector and mode resolved $\lambda_{\mathbf{q}\nu}$ varies dramatically with thickness. For instance, for the undoped bilayer 1T' MoTe$_2$, different from the monolayer case, the TA mode is activated along Γ-X/2 line and the LA mode is suppressed in the bilayer case (Figure 4c). To enable a meaningful comparison with the monolayers under an equivalent degree of doping, the number of doping electrons (0.4 *e*) per unit cells of the bilayer cases doubles those of the monolayer case. The difference of EPC strength between the monolayer 1T' MoTe$_2$ ($\lambda$ = 0.67) and that of the bilayer ($\lambda$ = 0.70) is found to be small. Under the electron doping, the EPC strength of ZA and LA mode of monolayer is significantly stronger than that of the bilayer, as shown in Figure 4b and d, respectively. For hole doping, the $\lambda$ increases from 0.16 (monolayer) to 0.45 (bilayer) with 0.4 *h* per cell, and the detailed $\lambda_{\mathbf{q}\nu}$ of the acoustic modes are sketched in Figure S11. Our calculation reveals that the EPC strength is significantly promoted in the reduced thickness only when they are electron-doped.

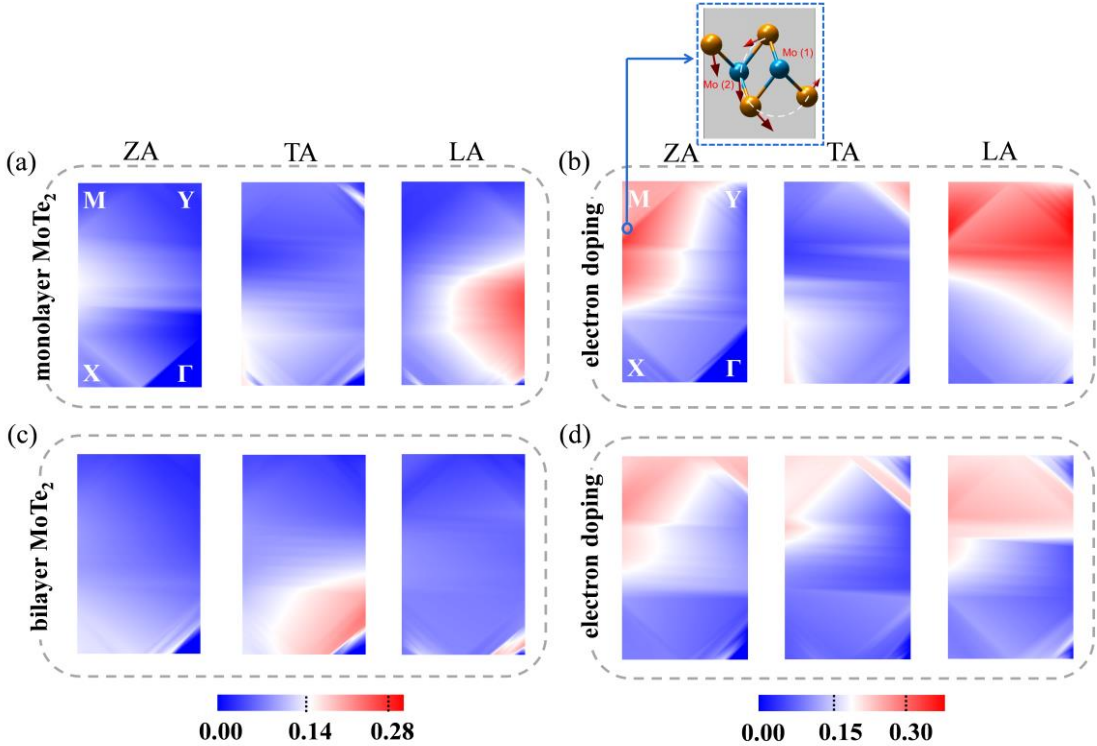

**Figure 4.** The evolution of wave vector and mode resolved $\lambda_{\mathbf{q}\nu}$ within the 2D Brillouin zone for acoustic modes of monolayer and bilayer MoTe$_2$ with different doping levels. (a) $n = 0$ and (b) $n_e = 9.19 \times 10^{13}$ e/cm$^2$ for 1T' monolayer. (c) $n = 0$ and (d) $n_e = 1.82$

× $10^{14}$ e/cm$^2$ for 1T' bilayer. The figure (a) and (c) adopt the same colorbar, while that of figure (b) and (d) adopt the same colorbar.

Accompanied with the varying EPC, we observe an increase of the $T_c$ in the case of electron doping, ranging from 0.67 K at the neutral sample to 4.13 K at 9.19× $10^{13}$ e/cm$^2$, and the value is 6 times larger than that of the pristine structure, indicating the $T_c$ is very sensitive to the electron doping. On the contrary, there is no phonon softening which is accompanied by sharp decreases of $T_c$ from 0.04 K at -4.73 × $10^{13}$ e/cm$^2$ to nearly 0 K at -9.55 × $10^{13}$ e/cm$^2$ under hole doping; these values are much lower than that of pristine monolayer 1T'-MoTe$_2$. Both parameter $\lambda$ and $N(E_f)$ also show the same tendency of the $T_c$, while the logarithmic average of frequencies ($\omega_{\log}$) decreases steadily with electron doping $n$, along with an insensitive behavior under hole doping (Figure S12). Here the $\lambda$ as an important indicator reflecting the strength of electron-phonon interactions is concomitant with the $T_c$ and can be promoted via electron doping.

Significantly, a strong renormalization of the $A_u$ mode (~180 cm$^{-1}$) is found at the X point. As shown in Figure 5, under electron doping, the mode shows a sharp softening, forming a dipped curve at the X, while a hardening occurs at the $\Gamma$ point. Under hole doping, a reversed trend is observed, and the $A_u$ branch gradually flattens along X-$\Gamma$. An increasing linewidth over the broader M-X-$\Gamma$ occurs simultaneously under electron doping. The existence of band nesting could account for the promoted EPC as we analyzed below.

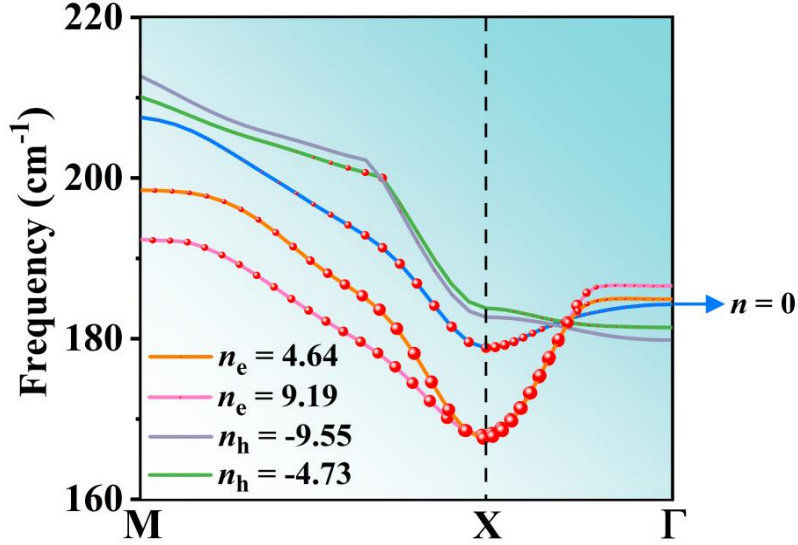

**Figure 5.** The evolution of phonon linewidth ($\gamma_{\mathbf{q}\nu}$) for infrared-active $A_u$ branch under different doping concentrations (unit of $10^{13}$ e/cm$^2$), where radius of red circles is proportional to phonon linewidths.

Under doping, the rigid-band approximation suggests the shift of the Fermi level to (de)populate these bands. The band containing the "H1" valley with a flatter and bonding nature induces a stronger density of states being sampled than the steeper band "H2". Bowing along Γ-Y and also Γ-Y', the electronic valleys renders intraband scattering via short-wavelength phonons within Γ-Y as required by momentum conservation. This explains the activation of EPC of the ZA (i.e. the circularly rocking mode in Figure 4b) and LA modes under electron doping. As shown in Figure 1a and b, around the Γ point the bands containing the "H1" and "H2" hole pockets cross Fermi level, making them semimetallic which is consistent with previous studies.[20] The two bands along X-Γ are overall parallel to each other with some shift in momentum, making them nesting and equispaced. Such nesting bands crossing the Fermi level enable the electronic coupling with broadly distributed phonons since the two bands are not perfectly equispaced. The coupling can be modulated by rigidly upward shifting of the Fermi level under different charge dopings. The 2D plot of the Fermi surface shows the presence of two separated electron pockets and one hole pocket centered on Γ-point in Figure 1d. Besides that, there is a conduction band minimum locate at Λ point along Γ to Y line, upon slightly doping electrons, the Fermi level shifts rigidly upwardly and enables the population of the second valley at X point, as shown in Figure S5. When both the Λ and X valleys are simultaneously populated, the strength of EPC in electron-doped MoTe$_2$ increases significantly, which is similar to doped MoS$_2$.[50,51] The corresponding 2D Fermi surfaces are given in Figure S6, we noted that the hole pocket at Γ point disappears and electron pocket emerges at X point in the case of *n*-type doping, while for hole doping the hole pocket at Γ point becomes larger due to a downward shift of Fermi level toward the deep valance band, along with the shrinkage of the former two electron pockets at X point. This is likely to account for the phonon softening of A$_u$ mode at X (Figure 5) and create additional channels to enable intra/inter-band electron-hole hopping via low-energy phonons, and accordingly the significantly promoted EPC for ZA and LA phonon branches as shown in Figure 4b. The activation of the low-energy acoustic modes under doping of electrons could be the reason of the enhanced $T_c$ temperature observed in experiments via doping.[35]

Importantly, consistent with the monolayer case, the acoustic phonons in bilayer (Figure S13) and bulk case (Figure S14) and bilayer T$_d$ phase (Figure S15) overall have a much stronger EPC than optical modes, as evidenced by the broadening mode resolved $\lambda_{\mathbf{q}\nu}$. Our results show the acoustic modes play a decisive role in modulate the superconductivity of MoTe$_2$, and charge doping can further modulate the EPC of these acoustic modes, i.e. a suppressed involvement of LA for scattering electrons under hole doping. This reasonably explains the experimental pressure-induced superconductivity[3,21] as the long-wavelength acoustic modes alter via shearing or compressive deformations under pressure.

## CONCLUSIONS

The magnitude of the EPC of semi-metallic monolayer MoTe$_2$, a prototype of the Weyl superconductor, is examined by first-principles calculations. Our calculation shows that the LA and ZA modes become activated with a promoted coupling strength under electron doping, while the TA modes largely keep silent. The presence of the nested equispaced conduction and valence bands around the Fermi level allows the sensitive modulation of the scattering channels with doping via low-energy acoustic phonons. Through examining the monolayer and bilayer MoTe$_2$, the EPC strength is found to be more sensitive to the charge doping than the thickness. Since the Fermi surface varies with the thickness and doping level, different phonon branches are activated for the charged doped monolayer and bilayer semimetallic MoTe$_2$, i.e. the ZA and LA modes playing a decisive role in the enhancement of EPC for electron-doped monolayer phase while the ZA and TA modes are responsible for the promoted EPC in electron-doped bilayer phase. These findings explain dramatically promoted superconducting critical temperature under chemical doping in experiments. The current work highlights the significance of precise activation of acoustic phonon for its role in coupling with modulated electronic structures under doping.

*Note added*: During the preparation of the response of the manuscript, there is a latest experimental demonstration[52] of superconductivity in bilayer T$_d$-MoTe$_2$ controlled by doping and polarization.

## AUTHOR INFORMATION

### Corresponding Authors

**Yongqing Cai** - *Joint Key Laboratory of the Ministry of Education, Institute of Applied Physics and Materials Engineering, University of Macau, Taipa, Macau, China;*
orcid.org/0000-0002-3565-574X;
Email: yongqingcai@um.edu.mo

### Authors

**Xiangyue Cui** - *Joint Key Laboratory of the Ministry of Education, Institute of Applied Physics and Materials Engineering, University of Macau, Taipa, Macau, China*

**Hejin Yan** - *Joint Key Laboratory of the Ministry of Education, Institute of Applied*

*Physics and Materials Engineering, University of Macau, Taipa, Macau, China*

**Xuefei Yan** - *School of Microelectronics Science and Technology, Sun Yat-Sen University, Zhuhai 519082, China; Guangdong Provincial Key Laboratory of Optoelectronic Information Processing Chips and Systems, Sun Yat-Sen University, Zhuhai 519082, China*

**Kun Zhou** - *School of Mechanical and Aerospace Engineering, Nanyang Technological University, 50 Nanyang Avenue, Singapore 639798, Singapore; Environmental Process Modelling Centre, Nanyang Environment and Water Research Institute, Nanyang Technological University, 1 Cleantech Loop, Singapore 637141, Singapore*

**Notes**

The authors declare no competing financial interest.

## ACKNOWLEDGEMENTS

This work was supported by the National Natural Science Foundation of China (Grant 22022309), and the Natural Science Foundation of Guangdong Province, China (2021A1515010024), the University of Macau (MYRG2020-00075-IAPME), the Science and Technology Development Fund from Macau SAR (FDCT-0163/2019/A3). This work was performed at the High Performance Computing Cluster (HPCC), which is supported by the Information and Communication Technology Office (ICTO) of the University of Macau.

## REFERENCE

(1) Yin, X. M.; Tang, C. S.; Zheng, Y.; Gao, J.; Wu, J.; Zhang, H.; Chhowalla, M.; Chen, W.; Wee. A. T. S. Recent Developments in 2D Transition Metal Dichalcogenides: Phase Transition and Applications of the (Quasi-) Metallic Phases. *Chem. Soc. Rev.* **2021**, *50*, 10087-10115.
(2) Pace, S.; Martini, L.; Convertino, D.; Keum, D. H.; Forti, S.; Pezzini, S.; Fabbri, F.; Mišeikis, V.; Coletti, C. Synthesis of Large-Scale Monolayer 1T'-MoTe$_2$ and Its Stabilization via Scalable hBN Encapsulation. *ACS Nano*. **2021**, *15*, 4213-4225.

(3) Qi, Y. P.; Naumov, P. G.; Ali, M. N.; Rajamathi, C. R.; Schnelle, W.; Barkalov, O.; Hanfland, M.; Wu, S.-C.; Shekhar, C.; Sun, Y.; Süβ, V.; Schmidt, M.; Schwarz, U.; Pippel, E.; Werner, P.; Hillebrand, R.; Förster, T.; Kampert, E.; Parkin, S.; Cava, R. J.; Felser, C.; Yan, B.; Medvedev, S. A. Superconductivity in Weyl Semimetal Candidate MoTe$_2$. *Nat. Commun.* **2016**, *7*, 11038.

(4) Mak, K. F.; Lee, C.; Hone, J.; Shan, J.; Heinz, T. F. Atomically Thin MoS$_2$: A New Direct-Gap Semiconductor. *Phys. Rev. Lett.* **2010**, *105*, 136805.

(5) Jariwala, D.; Sangwan, V. K.; Lauhon, L. J.; Marks, T. J.; Hersam, M. C. Emerging Device Applications for Semiconducting Two-Dimensional Transition Metal Dichalcogenides. *ACS Nano.* **2014**, *8*, 1102-1120.

(6) Radisavljevic, B.; Radenovic, A.; Brivio, J.; Giacometti, V.; Kis, A. Single-layer MoS$_2$ Transistors. *Nature Nanotech.* **2011**, *6*, 147-150.

(7) Lopez-Sanchez, O.; Lembke, D.; Kayci, M.; Radenovic, A.; Kis, A. Ultrasensitive Photodetectors Based on Monolayer MoS$_2$. *Nature Nanotech.* **2013**, *8*, 497-501.

(8) Xia, F.; Wang, H.; Xiao, D. Dubey, M.; Ramasubramaniam, A. Two-dimensional Material Nanophotonics. *Nature Photon.* **2014**, *8*, 899-907.

(9) Wang, Q. H.; Kalantar-Zadeh, K.; Kis, A.; Coleman, J. N.; Strano, M. S. Electronics and Optoelectronics of Two-dimensional Transition Metal Dichalcogenides. *Nature Nanotech.* **2012**, *7*, 699-712.

(10) Garcia, J. H.; Cummings, A. W.; Roche, S.; Spin Hall Effect and Weak Antilocalization in Graphene/Transition Metal Dichalcogenide Heterostructures. *Nano Lett.* **2017**, *17*, 5078-5083.

(11) Xu, X.; Yao, W.; Xiao, D.; Heinz, T. F. Spin and Pseudospins in Layered Transition Metal Dichalcogenides. *Nature Phys.* **2014**, *10*, 343-350.

(12) Zhou, L.; Xu, K.; Zubair, A.; Liao, A. D.; Fang, W.; Ouyang, F.; Lee, Y.-H.; Ueno, K.; Saito, R.; Palacios, T.; Kong, J.; Dresselhaus, M. S. Large-Area Synthesis of High-Quality Uniform Few-Layer MoTe$_2$. *J. Am. Chem. Soc.* **2015**, *137*, 11892-11895.

(13) Qian, X.; Liu, J.; Fu, L.; Li, J. Quantum Spin Hall Effect in Two-Dimensional Transition Metal Dichalcogenides. *Science* **2014**, *346*, 1344-1347.

(14) Deng, Y.; Zhao, X.; Zhu, C.; Li, P.; Duan, R.; Liu, G.; Liu, Z. MoTe$_2$: Semiconductor or Semimetal? *ACS Nano* **2021**, *15*, 12465-12474.

(15) Naylor, C. H.; Parkin, W. M.; Ping, J.; Gao, Z.; Zhou, Y. R.; Kim, Y.; Streller, F.; Carpick, R. W.; Rappe, A. M.; Drndić, M.; Kikkawa, J. M.; Charlie Johnson, A. T. Monolayer Single-Crystal 1T'-MoTe2 Grown by Chemical Vapor Deposition Exhibits Weak Antilocalization Effect. *Nano Lett.* **2016**, *16*, 4297-4304.

(16) Duerloo, K.-A. N.; Li, Y.; Reed, E. J. Structural Phase Transitions in Two-dimensional Mo- and W-dichalcogenide Monolayers. *Nat. Commun.* **2014**, *5*, 4214.

(17) Keum, D. H.; Cho, S.; Kim, J. H.; Choe, D.-H.; Sung, H.-J.; Kan, M.; Kang, H.; Hwang, J.-Y.; Kim, S. W.; Yang, H.; Chang, K. J.; Lee, Y. H. Bandgap Opening in Few-layered Monoclinic MoTe$_2$. *Nature Phys.* **2015**, *11*, 482-486.

(18) Yuan, J.; Chen,Y.; Xie, Y.; Zhang, X.; Rao, D.; Guo, Y.; Yan, X.; Feng, Y. P.; Cai, Y. Squeezed Metallic Droplet with Tunable Kubo Gap and Charge Injection in Transition Metal Dichalcogenides. *Proc. Natl. Acad. Sci. U. S. A.* **2020**, *117*, 6362-6369.

(19) Wang, W.; Kim, S.; Liu, M.; Cevallos, F. A.; Cava, R. J.; Ong, N. P. Evidence for An Edge Supercurrent in the Weyl Superconductor MoTe$_2$. *Science* **2020**, *368*, 534-537.

(20) Rhodes, D. A.; Jindal, A.; Yuan, N. F. Q.; Jun, Y.; Antony, A.; Wang, H.; Kim, B.; Chiu, Y.;

Taniguchi, T.; Watanabe, K.; Barmak, K.; Balicas, L.; Dean, C. R.; Qian, X.; Fu, L.; Pasupathy, A. N.; Hone, J. Enhanced Superconductivity in Monolayer $T_d$-MoTe$_2$. *Nano Lett*. **2021**, *21*, 2505-2511.

(21) Heikes, C.; Liu, I.-L.; Metz, T.; Eckberg, C.; Neves, P.; Wu, Y.; Hung, L.; Piccoli, P.; Cao, H.; Leao, J.; Paglione, J.; Yildirim, T.; Butch, N. P.; Ratcliff, W. Mechanical Control of Crystal Symmetry and Superconductivity in Weyl Semimetal MoTe$_2$. *Phys. Rev. Mater*. **2018**, *2*, 074202.

(22) Shi, W.; Ye, J.; Zhang, Y.; Suzuki, R.; Yoshida, M.; Miyazaki, J.; Inoue, N.; Saito, Y.; Iwasa, Y. Superconductivity Series in Transition Metal Dichalcogenides by Ionic Gating. *Sci. Rep*. **2015**, *5*, 12534.

(23) Si, C.; Liu, Z.; Duan, W.; Liu, F. First-principles Calculations on the Effect of Doping and Biaxial Tensile Strain on Electron-Phonon Coupling in Graphene. *Phys. Rev. Lett*. **2013**, *111*, 196802.

(24) Xi, X.; Wang, Z.; Zhao, W.; Park, J.-H.; Law, K. T.; Berger, H.; Forró, L.; Shan, J.; Mak, K. F. Ising Pairing in Superconducting NbSe$_2$ Atomic Layers. *Nature Phys*. **2016**, *12*, 139-143.

(25) Navarro-Moratalla, E.; Island, J. O.; Mañas-Valero, S.; Pinilla-Cienfuegos, E.; Castellanos-Gomez, A.; Quereda, J.; Rubio-Bollinger, G.; Chirolli, L.; Silva-Guillén, J. A.; Agraït,Paudyal, H.; Poncé, S.; Giustino, F.; Margine, E. R. Superconducting Properties of MoTe$_2$ from *ab initio* Anisotropic Migdal-Eliashberg Theory. *Phys. Rev. B* **2020**, *101*, 214515.

(26) Yang, W.; Mo, C.-J.; Fu, S.-B.; Yang, Y.; Zheng, F.-W.; Wang, X.-H.; Liu, Y.-A.; Hao, N.; Zhang, P. Soft-Mode-Phonon-Mediated Unconventional Superconductivity in Monolayer 1T'-WTe$_2$. *Phys. Rev. Lett*. **2020**, *125*, 237006.

(27) Sajadi, E.; Palomaki, T.; Fei, Z.; Zhao,W.; Bement, P.; Olsen, C.; Luescher, S.; Xu, X.; Folk, J. A.; Cobden, D. H. Gate-induced Superconductivity in A Monolayer Topological Insulator. *Science* **2018**, *362*, 922-925.

(28) Fatemi, V.; Wu, S.; Cao, Y.; Bretheau, L.; Gibson, Q. D.; Watanabe, K.; Taniguchi, T.; Cava, R. J.; Jarillo-Herrero, P. Electrically Tunable Low-density Superconductivity in A Monolayer Topological Insulator. *Science* **2018**, *362*, 926-929.

(29) Li, Y.; Gu, Q.; Chen, C.; Zhang, J.; Liu, Q.; Hu, X.; Liu, J.; Liu, Y.; Ling, L.; Tian, M.; Wang, Y.; Samarth, N.; Li, S.; Zhang, T.; Feng, J.; Wang, J. Nontrivial Superconductivity in Topological MoTe$_{2-x}$S$_x$ Crystals. *Proc. Natl. Acad. Sci. U. S. A*. **2018**, *115*, 9503-9508.

(30) Li, P.; Cui, J.; Zhou, J.; Guo, D.; Zhao, Z.; Yi, J.; Fan, J.; Ji, Z.; Jing, X.; Qu, F.; Yang, C.; Lu, L.; Lin, J.; Liu, Z.; Liu, G. Phase Transition and Superconductivity Enhancement in Se-Substituted MoTe$_2$ Thin Films. *Adv. Mater*. **2019**, *31*, 1904641.

(31) Takahashi, H.; Akiba, T.; Imura, K.; Shiino, T.; Deguchi, K.; Sato, N. K.; Sakai, H.; Bahramy, M. S.; Ishiwata, S. Anticorrelation Between Polar Lattice Instability and Superconductivity in the Weyl Semimetal Candidate MoTe$_2$. *Phys. Rev*. B **2017**, *95*, 100501(R).

(32) Mandal, M.; Marik, S.; Sajilesh, K. P.; Arushi; Singh, D.; Chakraborty, J.; Ganguli, N.; Singh, R. P. Enhancement of Superconducting Transition Temperature by Re Doping in Weyl Semimetal MoTe$_2$. *Phys. Rev. Mater*. **2018**, *2*, 094201.

(33) Lee, J.-H.; Son, Y.-W. Gate-tunable Superconductivity and Charge-density Wave in Monolayer 1T'-MoTe$_2$ and 1T'-WTe$_2$. *Phys. Chem. Chem. Phys*. **2021**, *23*, 17279-17286.

(34) Kim, D.; Lee, J.-H.; Kang, K.; Won, D.; Kwon, M.; Cho, S.; Son, Y.-W.; Yang, H. Thermomechanical Manipulation of Electric Transport in MoTe$_2$. *Adv. Electron Mater*. **2021**, *7*, 2000823.

(35) Cho, S.; Kang, S. H.; Yu, H. S.; Kim, H. W.; Ko, W.; Hwang, S. W.; Han, W. H.; Choe, D.-H.; Jung, Y. H.; Chang, K. J. Te-vacancy Driven Superconductivity in Orthorhombic Molybdenum Ditelluride. *2D Mater*. **2017**, *4*, 021030.

(36) Kim, H.-J.; Kang, S.-H.; Hamada, I.; Son, Y.-W. Origins of the structural phase transitions in MoTe$_2$ and WTe$_2$. *Phys. Rev. B* **2017**, *95*, 180101(R).

(37) Ma, T.; Chen, H.; Yananose, K.; Zhou, X.; Wang, L.; Li, R.; Zhu, Z.; Wu, Z.; Xu, Q.-H; Yu, j.; Qiu, C. W.; Stroppa, A.; Loh, K. P. Growth of bilayer MoTe$_2$ single crystals with strong non-linear Hall effect. *Nat. Commun.* **2022**, *13*, 5465.

(38) Perdew, J. P.; Burke, K.; Ernzerhof, M. Generalized Gradient Approximation Made Simple. *Phys. Rev. Lett.* **1996**, *77*, 3865.

(39) Giannozzi, P.; Baroni, S.; Bonini, N.; Calandra, M.; Car, R.; Cavazzoni, C.; Ceresoli, D.; Chiarotti, G. L.; Cococcioni, M.; Dabo, I.; Corso, A. Dal; Fabris, S.; *et al*., *J. Phys.: Condens. Matter* QUANTUM ESPRESSO: A Modular and Open-source Software Project for Quantum Simulations of Materials. **2009**, *21*, 395502.

(40) Hu, H.; Liu, M.; Wang, Z. F.; Zhu, J.; Wu, D.; Ding, H.; Liu, Z.; Liu, F. Quantum Electronic Stress: Density-Functional-Theory Formulation and Physical Manifestation. *Phys. Rev. Lett*. **2012**, *109*, 055501.

(41) Baroni, S.; de Gironcoli, S.; Corso, A. D.; Giannozzi, P. Phonons and Related Crystal Properties from Density-functional Perturbation Theory. *Rev. Mod. Phys.* **2001**, *73*, 515.

(42) Paudyal, H.; Poncé, S.; Giustino, F.; Margine, E. R. Superconducting Properties of MoTe$_2$ from *ab initio* Anisotropic Migdal-Eliashberg Theory. *Phys. Rev. B* **2020**, *101*, 214515.

(43) Kresse, G.; Furthmüller, J. Efficient Iterative Schemes for *ab* initio Total-energy Calculations Using A Plane-wave Basis Set. *Phys. Rev. B* **1996**, *54*, 11169.

(44) Kan, M.; Nam, H. G.; Lee, Y. H.; Sun, Q. Phase Stability and Raman Vibration of the Molybdenum Ditelluride (MoTe$_2$) Monolayer. *Phys. Chem. Chem. Phys*. **2015**, *17*, 14866-14871.

(45) Allen, P. B.; Dynes, R. C. Transition temperature of strong-coupled superconductors reanalyzed. *Phys. Rev. B* **1975**, *12*, 905.

(46) Bardeen, J.; Cooper, L. N.; Schrieffer, J. R. Theory of Superconductivity. *Phys. Rev*. **1957**, *108*, 1175.

(47) Allen, P. B. Neutron Spectroscopy of Superconductors. *Phys. Rev. B* **1972**, *6*, 2577.

(48) Giustino, F. Electron-phonon Interactions from First Principles. *Rev. Mod. Phys*. **2017**, *89*, 015003.

(49) Gao, M.; Li, Q.-Z.; Yan, X.-W. Wang, J. Prediction of Phonon-mediated Superconductivity in Borophene. *Phys. Rev. B* **2017**, *95*, 024505.

(50) Ge. Y.; Liu, A. Y. Phonon-mediated superconductivity in electron-doped single-layer MoS$_2$: A first-principles prediction. *Phys. Rev. B* **2013**, 87, 241408.

(51) Sohier, T.; Ponomarev, E.; Gibertini, M.; Berger, H.; Marzari, N.; Ubrig, N.; Morpurgo, A.F. Enhanced Electron-Phonon Interaction in Multivalley Materials. *Phys. Rev. X* **2019**, *9*, 031019.

(52) Jindal, A.; Saha, A.; Li, Z.; Taniguchi, T.; Watanabe, K.; Hone, J. C.; Birol, T.; Fernandes, R. M.; Dean, C. R.; Pasupathy, A. N.; Rhodes D. A. Coupled ferroelectricity and superconductivity in bilayer T$_d$-MoTe$_2$. *Nature* **2023**, 613, 48.